# Cultural Astronomy and Archaeoastronomy: an Italian Experience[1]

**Elio Antonello**
INAF – Osservatorio Astronomico di Brera
Via E. Bianchi 46, 23807 Merate, Italy
elio.antonello@brera.inaf.it

**Abstract**
A brief review is given of some recent positive developments regarding the reception of archaeoastronomy by the archaeological institutions in Italy. Discussions and problems that are currently going on in this field are also mentioned, such as the separation of the scientific and humanistic disciplines (i.e. 'the two cultures' problem). Suggestions based on contemporary philosophy are also reported. Finally, 'sky-gazing' is proposed as the place where the two cultures could meet, since, taking Plato into account, sky-gazing could be considered the 'mom' of the human knowledge, and of the scientific and humanistic disciplines.

**Archaeoastronomy and archaeologists**

Some years ago, several fundamental problems were pointed out at conferences held at the Accademia Nazionale dei Lincei in Italy (in 1994, 1997 and 2000) which were dedicated to archaeoastronomy and cultural astronomy. These problems are well known, since they are common to our field. Let me begin by discussing the lack of interest shown by archaeologists. In 1997 (and 2000), Gustavo Traversari, one of the few archaeologists who was already very interested in archaeoastronomy, lamented this situation: 'We see that at this Conference archaeologists are almost completely absent. Perhaps they are not interested. I know professors who say that archaeoastronomy is a strange and incomprehensible discipline' (Traversari 1998: 254; 2000: 405). It seems that presently in Italy we are slowly overcoming this problem, and Traversari's view may appear a bit too pessimistic. Before the Italian Society for Archaeoastronomy (SIA) was founded, in the year 2000, three conferences had been organized at the Accademia Nazionale dei Lincei, and after its founding the SIA met once a year with an increasing presence of archaeologists at its meetings. Our last annual conference was held in 2011 in Bologna and also in the Etruscan town of Misa near Marzabotto. It was organized in conjunction with both of the astronomical and archaeological institutions of Bologna, and at the conference the archaeologists in attendance had the opportunity to expound their theories and views of our field.

Another encouraging piece of news is that SIA is a member of the Italian Institute of Prehistory and Protohistory. This Institute brings together archaeological institutions such as Departments of Archaeology, Superintendences and Museums, and its purpose is to coordinate and promote scientific activity in relation to the study of prehistoric and proto-historic civilizations. Just as an example, in June, 2012, I attended a round table dedicated to the problems confronted by scientific disciplines that contribute to archaeological research (such as archaeobotanics, archaeozoology, dendrochronology, and geoarchaeology). It is understandable that, given the present lack of funds for the Italian culture, the round table was mainly a *cahier de doléances*.

---

## Separation of disciplines

Let me cite a second long-standing problem: the separation of disciplines. In 2000, the astrophysicist Vittorio Castellani said: 'Please, let me provoke you. I will tell you what interdisciplinarity means in Italy: people meet around a table that have different cultures, different languages, they talk to each other without understanding, then they get up and say: it was an excellent interdisciplinary meeting' (Castellani 2000: 406). Last year in Bologna archaeologists recognized the existence of the problem posed by the fact that each discipline has its own 'arrogance'. Every scholar thinks: 'My own discipline is more important than yours'. There is also the fear of not being properly respected by scholars working in other fields.

From the round-table discussion that took place during the SEAC meeting last year in Evora, I got just such a feeling: European archaeoastronomers are afraid; they are afraid to be considered nothing more than data suppliers by archaeologists. In Italy we are trying to overcome this problem by helping out archaeologists and anthropologists with their investigations, since they are often in need of some help in the interpretation of the data. There is a reason for archaeoastronomers to be, so to speak, altruist, unselfish, and I will try to explain why in Section Sky-gazing.

Let me remind you of the essay 'The two cultures' by Charles P. Snow, by quoting a columnist writing in the *Scientific American*. The summer of 2009 'marked the 50$^{th}$ anniversary of the famous essay, in which Snow lamented the great cultural divide that separates two great areas of human intellectual activity, science and the arts. Snow argued that practitioners in both areas should build bridges, to further the progress of human knowledge and to benefit society. Alas, Snow's vision has gone unrealized. […] Many of those in the humanities, arts and politics remain content living within the walls of scientific illiteracy' (Krauss 2009). Negative conclusions were drawn also by other scholars who commented the same essay (e.g. Jardine 2009).

Snow (1961: 17) claimed: 'There seems then to be no place where the cultures meet. I am not going to waste time saying that this is a pity. It is much worse than that'. Is there really no place?

The impressive sight of the starry night sky, far from light pollution, is so breath-taking that one cannot help but draw the following conclusion: there is plenty of 'space' out there. Some time ago, the archaeologist Tiziano Mannoni (2009: 52) said:

> I spent many nights outdoors during the [second world] war, in places with no lights at all, and when seeing all the nights what it was visible, one couldn't help considering and studying the sky. The link with the sky is completely different from that with the earth. There is not a 'physical' relation, but only a visual […] and mental one, that triggers the imagination and thoughts.

Similar statements regarding the differences between the kind of relation that humans have with the sky as opposed to the earth, were already forthcoming from archaeoastronomers. But Mannoni was an archaeologist. When I have had occasion to talk with archaeologists, they have expressed similar views. We have such thoughts because before being astronomers or archaeologists, we are just human beings. Hence, perhaps sky-gazing (or cultural astronomy) could be the place where it is possible to establish a dialogue between the two cultures, since both humanists and scientists are saying the same thing. This point will be discussed further in Section Sky-gazing.

## Language and mentality

### Western culture

The archaeologist Giovanni Lilliu (1998: 252) strongly criticized some Sardinian amateurs when they used the expression 'astronomical observatory' to refer to a *nuraghe*. He specified that a

*nuraghe* was actually just a sort of castle, with some military equipment. We could imagine, however, there were sentinels for night watching, and may be someone was charged also with sky-gazing. The question is what terminology should be used in such a case. This is a very general problem, and also one that exists in archaeology, I guess, since it concerns the terms that should be used when describing the distant past. The term 'astronomy' itself, for instance, is misleading, since astronomy should indicate the scientific study of the sky, understood in a rather modern sense; therefore, when talking about the distant past (i.e. prehistory), sky-gazing would appear more appropriate than astronomy. Sometimes we look far into the past in a way similar to that represented in 'The Flintstones' movie; even if we pay much attention, we cannot avoid to projecting our own world-view back in time.

An analogy is afforded by the ethnological studies of modern-day 'primitive' populations, such as those carried out by Levy Bruhl (1922). It is difficult to translate faithfully the language of the 'primitives' because it requires being able to interpret their thought processes in a way that is consonant with the indigenous cultural frames of reference. And, of course, when attempting to produce reliable descriptions, one is confronted with the problem of having to utilize terms drawn from Western culture, words and expression that are already charged with highly specific meanings, that do not necessarily reflect the cultural conceptualizations native to the frames of reference of the peoples being described. Sometimes even the *logic* of such 'primitive populations' appears to be different from the Aristotelian one touted by Western culture. In other words, it is difficult to understand the 'primitive mentality' since we are studying it from within our Western cultural system. Consequently, there is always the risk of assuming a colonialist-like attitude, by imposing in some way our Western mentality on non-Western populations. In a similar way, when we study the peoples of the past we run the risk of imposing our modern scientific mentality on them. In particular, that may occur when archaeoastronomers try to provide proofs for their own hypotheses concerning the astronomical content of archaeological finds and sites, instead of following the archaeology tenet: let the stones speak for themselves.

The Western mentality seems to be based on a Heraclitean vision of the world ('everything flows'), with absolute space and time as a background. A few centuries ago, Newton assumed time as absolute, well aware of the fact that it was an assumption that couldn't be proved. Today we are educated in the practical every day experience where 'becoming' and absolute space and time are self-evident (the theory of general relativity has not yet entered, so to speak, our everyday life, apart from the GPS system time corrections). To doubt the existence of 'becoming' would be just as absurd as to deny reality. However, those are only assumptions. In the next Subsection we will mention some philosophical discussions about these issues.

*A possible help from philosophy*

The archaeologist Julian Thomas discussed the possibility of a phenomenological archaeology, beyond the modernist approach that involves the radical separation of culture, nature, mind, body, society, individuals and artefacts:

> I will hope to show that cultural significance and the production of meaning are not encapsulated in any one sutured entity, whether mind, body, society or nature. Identity and meaning are both relational constructions, which emerge through the process of human Being-in-the-world. (Thomas 1996: 30)

With that purpose in mind, he considered the concept of temporality, drawing on Heidegger's philosophy. In such a context he therefore mentioned the phenomenon of 'Being' and 'the way that it has been covered over as a problem since the time of the Greeks'. He used Heidegger's term *Dasein*, which could indicate 'the way the human beings are: Dasein is a way of being' (Thomas 1996: 40).

> World is that which Dasein can cope with. As Dasein carries out its everyday activities, the things of the world show up to it as unconcealed […] The event of things showing up in this way can be termed as 'clearing'. Human beings, by being linguistic creatures, are engaged in a complex network of relationships, in which the material and the symbolic cannot be disentangled. Their grasp of this network at any given time is limited and regionalised, since it spreads out from a located spatial centre. This is the clearing. Dasein *is* the clearing, in that it is not a thing in itself: Dasein itself is a space or process in which things can be made to show up. (Thomas 1996: 69)

According to many scholars, Emanuele Severino is one of the most important Italian philosophers. I will try to recall some of his ideas, as far as I understood them, since they may be of some help in this context. He proposed a different ontology from Heidegger's. He emphasized one of the major conundrums of Western philosophy, that of the Greek sense of becoming: a being-in-the-world is a 'thing' coming from nothing and going back to nothing. He argued (with strong philosophical arguments) that is a 'folly', since every 'being' is 'eternal'. The 'becoming' is the 'folly' of the Western world, because this world has identified 'being' with 'nothing' (e.g. Severino 1995; unfortunately, there are few translations of his books). Instead of the 'clearing' discussed by Thomas, he has proposed a sort of 'circle of the coming into sight' (*cerchio dell'apparire*) in which the beings get in (become 'visible' as being-in-the-world) and from where they get out (not more 'visible'); however, the beings never are nothing, they are eternal. This picture may sound familiar to physicists since it reminds of the four-dimensional space-time of special relativity. There are seemingly some connections with Anglo-American eternalism (for a philosophical discussion of 'time', see e.g. Turetzky 1998); however, the 'circle' of Severino was not inspired by physics, since it would be actually an unavoidable consequence of the deep (abysmal) philosophical meaning of 'being'.

Severino suggests going back to Parmenides, and then finding another way, different from that of 'becoming' followed by Greek (and then by all the Western world) philosophers. Before that happens, however, it is necessary that the present world dominated by the 'technic' (or technology; *tecnica*) go ahead, along its (long) way until waning (*tramonto*). The technic should not be understood in a negative sense at all. Its undisputed domain is an unavoidable consequence of modern nihilism, the extreme outcome of 'becoming', since in the last centuries all higher metaphysical absolute values, such as 'God' and 'truth', were not able to resist the philosophical criticism levelled at them and succumbed (note that in this Subsection we are dealing essentially with philosophical issues).

It seems to me that the big problems mentioned in the present paper, such as the separation of disciplines, the limits of the western mentality, and also the radical separation remarked by Thomas, are strictly related to our way of thinking based on the Greek sense of 'becoming'. The idea that 'things' come from 'nothing' and go back to 'nothing' permeates all of Western culture. More generally, 'things' can be (and therefore are) 'nothing', i.e. they can be ignored, excluded or nullified; this is a necessary approach, for example, in the scientific method and reductionism. Probably we should summarize the problematic situation we are experiencing as the conflict between 'separation' and 'relation'. Maybe Severino's ideas could be of help in this context, since they would allow us to see the world under a different light, though the way they could help effectively is a program whose feasibility has yet to be demonstrated.

**Sky-gazing**

Often it is possible to read sentences where mathematics, or philosophy, is defined as the mother of sciences, or the mother of knowledge. Nevertheless, it seems to me that 'sky-gazing' could be considered the 'mom' of human reflection, of human knowledge, and of all the disciplines, including mathematics and philosophy, since it triggers imagination and thoughts, as suggested by astronomers and archaeologists. Note that I mean mom rather than mother, and, of course, I mean a standard (good) mom. Moreover, note that mom's house is the place where 'children' now adults

(the disciplines) can convene, leaving outdoor arrogance (at least, usually it is left outdoors). Cultural 'astronomy', in a broad sense, could be such a house, since it deals with sky-gazing (of every epoch), and therefore it could be the place where the 'two cultures' could meet.

It may appear that I appeal to emotional rather than rational intelligence; note however that it is intelligence anyway. In any case, apart from the word 'mom', what I am saying is not new. Plato wrote:

> Vision, in my view, is the cause of the greatest benefit to us, inasmuch as none of the accounts now given concerning the Universe would ever have been given if men had not seen the stars or the sun or the heaven. But as it is, the vision of day and night and of months and circling years has created the art of number […] and has given us not only the notion of Time but also means of research into the nature of the Universe. From these we have procured philosophy […] in all its range. (Plato, *Timaeus* [47a]; translation: W.R.M. Lamb).

We might conclude, therefore, that sky-gazing is the mom of mathematics, 'the art of number', and philosophy, 'in all its range'. Further interesting concepts are expressed in *Epinomis*:

> […] for why should we not believe the cause of all the good things that are ours to have been the cause also of what is far the greatest, understanding? And who is it that I magnify with the name of God? Merely Heaven [...] That it has been the cause of all the other good things we have, we shall all admit; that it likewise gave us number we do really say. […] In bespangling itself and turning the stars that it contains, it produces all their courses and the seasons and food for all […] And thence, accordingly, we have understanding in general, we may say, and therewith all number, and all other good things […] Now let us go on to inquire into the question of how we learnt to count in numbers. Tell me, whence have we got the conception of one and two […]? Among such things, what one more singularly beautiful can a man behold than the world of day? Then he comes to the province of night, and views it; and there quite another sight lies before him. And so the heaven, revolving these very objects for many nights and many days, never ceases to teach men one and two, until even the most unintelligent have learnt sufficiently to number; for that there are also three and four and many. (Plato's school, *Epinomis* [976e; 977a, b; 978b, c, d]; translation W.R.M. Lamb)

Heaven should have a lot of patience, such as a mom with her child, if it has to wait even for the most unintelligent man.

Once again, the plain conclusion would be that sky-gazing is the mom of human knowledge and then of the disciplines: they are her daughters. But what about archaeoastronomy? Is it a discipline? In my opinion, it is not a subdiscipline of archaeology as was stated by the archaeologist Colin Renfrew (2007: 184). In my opinion it would not be even a scientific discipline; certainly it is not hard science. It is a set of astronomical, mathematical and statistical methods lent to archaeology and anthropology, and these scientific methods must be applied rigorously, on a case by case basis. Therefore, archaeoastronomy helps or serves archaeology; however, it is not at all a servant of archaeology. Archaeoastronomy is a practical manifestation or, so to speak, a practical expression of 'mom' who operates through archaeoastronomers. Everybody knows that mom is not a servant, that she is altruistic and asks nothing in return. She helps in this way her adult children, the various disciplines, even today.

As far as I know, up to now nobody has succeeded in defining archaeoastronomy as a scientific discipline, with its own tenets and principles. I suspect that such an attempt would be nonsense. Why nonsense? Because mom can't be the daughter of herself.

# Conclusion

I do not pretend to convince people here. I mentioned long-standing, secular problems that are deeply rooted in our world, and pretending to propose credible general solutions to them would be just foolish. What I tried to do has been to offer some reflections, with the hope they could be of some help.